\documentclass[aps,prb,twocolumn,showpacs,superscriptaddress,groupedaddress]{revtex4} 
\pdfoutput=1
\usepackage{graphicx}  
\usepackage{dcolumn}   
\usepackage{bm}        
\usepackage{amssymb}   
\usepackage{float}

\usepackage{amsmath,mathrsfs}
\usepackage{amsthm}
\usepackage{epsfig}

\usepackage{color}
\usepackage{soul}

\def\bra#1{\left\langle#1\right|}
\def\ket#1{\left|#1\right\rangle}

\begin{document}

\title{Position-momentum Duality in the Entanglement Spectrum of Free Fermions}

\author{Ching Hua Lee}
\affiliation{Department of Physics, Stanford University, Stanford, CA 94305, USA}
\author{Peng Ye}
\affiliation{Perimeter Institute for Theoretical Physics, Waterloo, Ontario, Canada N2L 2Y5}
\author{Xiao-Liang Qi}
\affiliation{Department of Physics, Stanford University, Stanford, CA 94305, USA}

\date{\today}

\begin{abstract}
We propose an exact equivalence between the entanglement spectra of two completely different free-fermion systems at zero temperature. This equivalence follows from a position-momentum duality where the physical roles of the occupied band and real space projectors are exchanged. We examine the physical consequences of this duality in multi-band models, and as an example also physically motivate the equivalence of the entanglement spectrum of a real space partitioned two-band topological insulator with that of a bilayer Fermi gas with an interlayer partition. This duality has an interesting relation with the Wannier Spectrum in the high-temperature limit, and can also be extended to other basis-independent physical quantities like particle-number fluctuations. 
\end{abstract}

\maketitle

\section{Introduction}

Entanglement is a purely quantum phenomenon that distinguishes quantum systems from classical ones. Many measures have been proposed to characterize entanglement, among which the simplest is the von Neumann entanglement entropy (EE) which measures the bipartite entanglement between two subsystems while the whole system is in a pure state. The EE is determined from the reduced density matrix (RDM) $\rho$ via $S=-{\rm tr}\left(\rho\log\rho\right)$, whose spectrum known as the entanglement spectrum (ES) provides a more precise characterization of entanglement\cite{li2008}.

Recently, the EE and ES have been used to probe novel physical characteristics of quantum states in condensed matter. For example, the long distance behavior of the EE characterizes the central charge of a $(1+1)$-d conformal field theory\cite{holzhey1994}; a universal subleading constant term in the EE, known as the topological EE , probes the total quantum dimension of 2-d topologically ordered states\cite{kitaev2006,levin2006}; the Fermi gas in generic spatial dimensions $d>1$ and surface area $A$ is characterized by a super-area-law\cite{ryu2006,wolf2006,verstraete2006,plenio2005,gioev2006} with $S\propto A\log A$; in the presence of interactions, a similar logarithmic enhancement was suggested in Ref. \onlinecite{swingle2010entanglement} and explicitly derived in Ref. \onlinecite{ding2012} for fermions, and in Ref. \onlinecite{lai2013} for bosons; in various topological states, the ES has been shown to be topologically equivalent to the edge/surface states along a physical boundary\cite{li2008,regnault2009,laeuchli2010,yao2010,turner2010,fidkowski2010,chandran2011,qi2012}. The EE and ES also play essential role in the numerical method of Density Matrix Renormalization Group (DMRG)\cite{white1992}.

So far, most research on the EE and ES focuses on the characterization of a single state of matter. In this paper, we shall instead introduce an exact relation between the entanglement properties of two distinct states of matter. We focus on free fermion systems at zero temperature, and derive a simple duality that allows us to find two completely different systems with different real space partitions possessing identical ES.

As we will discuss in detail, the free fermion ES is determined by two projectors, one determined by the Fermi function at zero temperature, and the other by the real space partition. The two projectors play symmetric role in determining the ES, such that a ``dual" system can be defined by exchanging their roles. For translation-invariant systems, exchanging the two projectors entails re-interpreting real space and momentum space, hence making this mapping a position-momentum duality. Such a symmetry between the two projectors have been discussed in previous works\cite{huang2012edge,huang2012,gioev2006}, but the resulting duality between different systems have not been studied.

In the following, we shall first provide a general derivation of the duality, and then discuss its application to multi-band models. 
A real space partitioned two-band topological insulator is dual to a bilayer Fermi gas with an interlayer partition. The former possesses a momentum space gauge field which is mapped to a real space non-Abelian gauge field in the Fermi gas. With such qualitatively different states, it is counter-intuitive that they have identical ES. Interestingly, the ES of the Fermi gas contains gapless modes similar to the topological edge states of the topological insulator, and we attempt to give an intuitive physical argument for that. This duality can also be extended to other physical quantities such as particle number fluctuations that can be probed in quantum noise measurements, Finally, we briefly explore the interesting relation between finite-temperature ES and the Wannier spectrum.

\section{General formula of the position-momentum duality}\label{sec:general}

Consider a free-fermion system with the Hamiltonian $H=\sum_{i,j}f_i^\dagger h_{ij}f_j$, with $f_i$ annihilation a fermion at site $i$. A real-space partition is defined by a subregion $A$ and its complement $B=\bar{A}$ in the system. Since all multi-point correlation functions obey Wick's theorem, the RDM $\rho_A$ for ground state $\ket{G}$ takes the Gaussian form\cite{peschel2002}
\begin{eqnarray}
\rho_A={\rm tr}\left[\ket{G}\bra{G}\right]=e^{-H_E},~H_E=\sum_{i,j\in A}f_i^\dagger {h_E}_{ij}f_j
\end{eqnarray}
The single-particle ``entanglement Hamiltonian" $h_E$ can be determined from the two-point correlation function ${C}_{ij}=\bra{G}f_i^\dagger f_j\ket{G},~i,j\in A$ via
\begin{equation}
h_E=\log\left(C^{-1}-\mathbb{I}\right)
\label{peschel}
\end{equation}
with $\mathbb{I}$ the identity matrix. Therefore the ES  of $\rho_A$, {\it i.e.} the spectrum of $H_E$, is determined by that of $h_E$ which is in turn determined by that of $C$.

$C$ is obtained by projecting the correlation matrix of the whole system onto the subsystem $A$. Defining $R=\sum_{i\in A}\ket{i}\bra{i}$ as the projection operator\cite{huang2012edge,huang2012,alexandradinata2011} onto $A$, one can write\footnote{To be more precise, $RPR$ is a matrix with the dimension of the total number of sites, while $C$ is its truncation to the $A$ subsystem. Since they only differ by trivial zero eigenvalues in the $B$ subsystem, we will simply identify $C$ with $RPR$ in the following.}
\begin{equation}
C= RPR
\label{rpr}
 \end{equation}
$P$ is also a projection operator, projecting to the occupied states via $P=\theta(-h)=\sum_n\theta(-\lambda_n)\ket{n}\bra{n}$. Here $\ket{n}$ and $\lambda_n$ are the eigenstates and eigenvalues of the single particle Hamiltonian $h$, and $\theta(x)$ is the step function. In general, $P$ and $R$ do not commute. For example, $P$ and $R$ are respectively the momentum and real space projections in a translationally-invariant Fermi gas.

The key mathematical reason behind the duality is that the eigenvalues of $C=RPR$ are identical to that of another operator $C'=PRP$ as long as both $P$ and $R$ are projectors, i.e. $P^2=P$ and $R^2=R$. This is easily shown as follows\cite{huang2012edge}: Suppose $C|\psi\rangle$ is an eigenstate of $C$ with eigenvalue $c$. Then $c |\psi\rangle = C|\psi\rangle =R^2PR|\psi\rangle=R(RPR)|\psi\rangle=c (R|\psi\rangle)$, so $R|\psi\rangle = |\psi\rangle $. $P|\psi\rangle$ will be an eigenstate of $C'=PRP$ with the same eigenvalue $c$, because $PRP(P|\psi\rangle)=PRP|\psi\rangle=PRP(R|\psi\rangle)=c(P|\psi\rangle)$. 
Denoting the eigenvalue spectrum of a matrix $C$ as ${\rm Spec}(C)$, we thus have
\begin{eqnarray}
{\rm Spec}(RPR)={\rm Spec(PRP)}\label{duality}
\end{eqnarray}
if $P$ and $R$ are both projectors. Now it is straightforward to define two free-fermion systems with identical ES. From a system with $P$ and $R$ projectors defined above, one can define a ``dual system" with $R\leftrightarrow P$ exchanged, i.e. $R$ being the projector to occupied states and $P$ the projector that defines the partition. Since the ES only depends on the eigenvalues of $C$ via Eq. \ref{peschel}, the two systems with correlation matrices $C=RPR$ and $C'=PRP$ have identical ES . The roles of the real-space basis $\ket{i}$ and the energy eigenstate basis $\ket{n}$ are also exchanged. In the dual system, $\ket{n}$ becomes the real-space sites, since $P$ is diagonal in this basis. When $\ket{i}$ lives on a lattice, so do ${\bf k}$. Consequently, the roles of the reciprocal and real space lattices are exchanged in the duality. In this sense, the $R\leftrightarrow P$ duality is indeed a position-momentum duality. 

While we have so far referred to $i$ as real-space 'sites', they can in general contain internal degrees of freedom (DOFs) like bands or spins/sublattices. We also emphasize that the duality still applies to systems without translational symmetry, since $P$ and $R$ are still well-defined.

\begin{figure}[H]
\begin{minipage}{0.99\linewidth}
\includegraphics[scale=0.31]{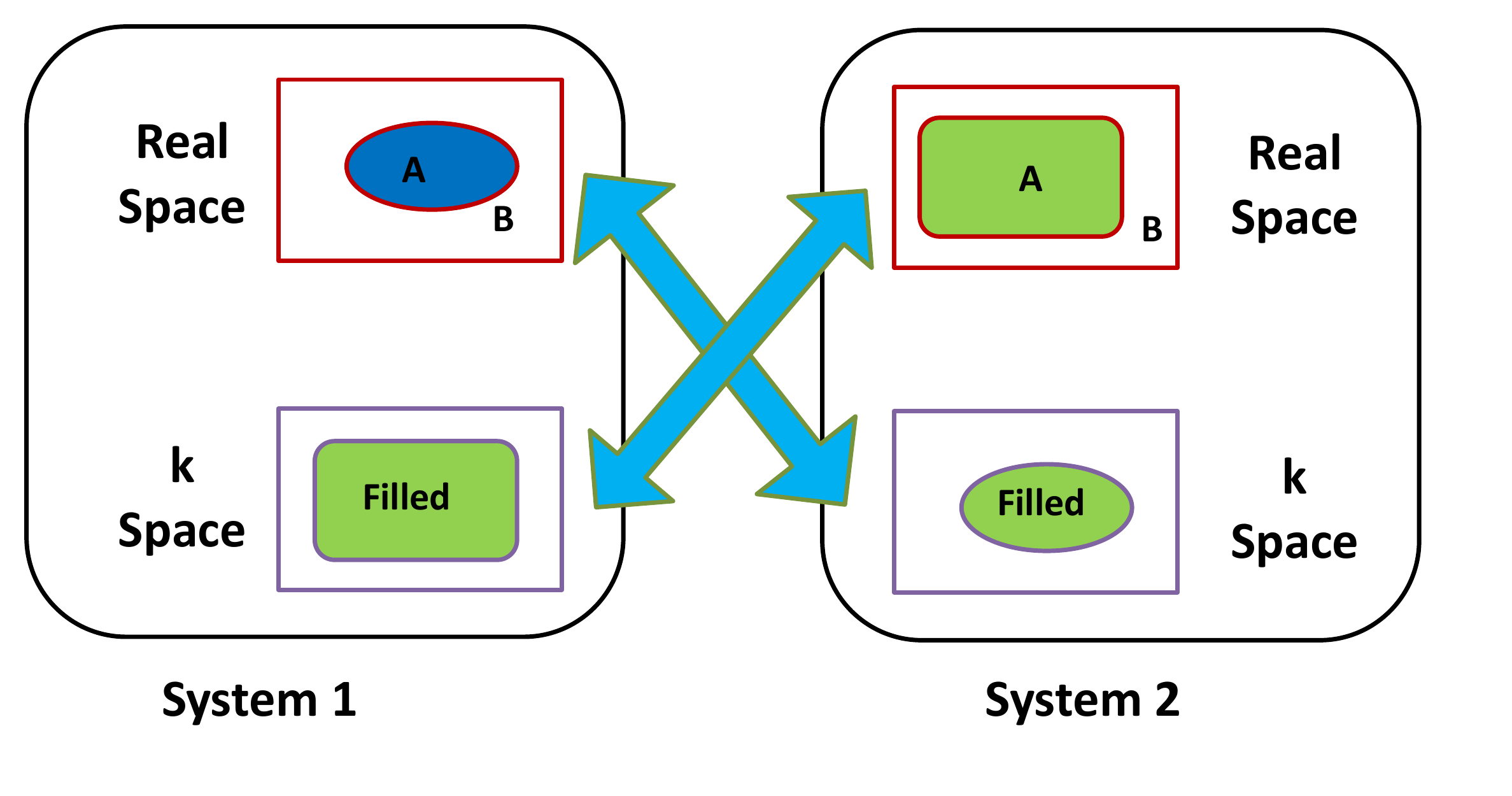}
\end{minipage}
\caption{(Color Online) Momentum space and real space are interchanged, i.e. an entanglement cut is mapped to a Fermi surface and vice versa.
This results in two dual systems with identical entanglement spectra. }\label{fig:duality}\end{figure}

\section{Examples of the duality}

\subsection{EE of a single-band Fermi gas}

As the simplest example, we consider a translationally invariant Fermi gas with only one energy band described by $H=\sum_{\bf k}\epsilon_{\bf k}f_{\bf k}^\dagger f_{\bf k}$. The projector $P=\sum_{{\bf k}\in \Gamma}\ket{\bf k}\bra{\bf k}$ acts in momentum space and projects onto $\Gamma$, the Fermi sea region where $\epsilon_{\bf k}<0$. The real space partition gives $R=\sum_{i\in A}\ket{i}\bra{i}$. Consider a periodic system in $d$-dimensions, so that both the real and reciprocal lattice are defined on a $d$-dimensional torus. When $R$ and $P$ are interchanged, the resulting dual system is still a single-band Fermi gas. But the momentum space region $\Gamma$ and the real space region $A$ will be exchanged, as is illustrated in Fig. \ref{fig:duality}. Therefore, we conclude that the real space and momentum space regions $A$ and $\Gamma$ must enter the expression for the EE and ES symmetrically. This is indeed consistent with
the rigorous result in Refs. \onlinecite{gioev2006,barthel2006,li2006,swingle2010} which was inspired by the Widom Conjecture\cite{widom1982}:
\begin{equation}
S_A\propto L^{d-1}\log L \int_{\partial \Gamma}\int_{\partial A}|n_x\cdot n_p|dS_x dS_p
\label{snp}\end{equation}
where $L$ sets the scale of the system size, $d$ is its spatial dimension, $\partial A$ and $\partial \Gamma$ are respectively the real space entanglement cut and the Fermi surface, and $n_x dS_x$, $n_p dS_p$ are their corresponding surface area elements.


The position-momentum duality also applies to the logarithmic part of the EE, beyond what is contained in Eq. \ref{snp}. Consider a 1-D periodic system of circumference $L$ with a spatial partition $A$ of length $L_A$. Suppose it also has $L_F$ of its $L$ momentum eigenstates occupied, corresponding, for instance, to Fermi points at $\pm\pi L_F/L$. Since it is a critical system, CFT gives us\cite{calabrese2004,calabrese2006} $S_A=\frac{1}{3}\log(L_{eff}\sin(\pi L_A/L))$, where $L_{eff}$ is proportional to $L$ and an unknown factor depending on the size of the Fermi sea. Upon interchanging the roles of position and momentum, we can also write $S_A=\frac{1}{3}\log(L'_{eff}\sin(\pi L_F/L))$. Comparing $L_{eff}$ with $L'_{eff}$, 
we obtain the symmetric expression

\begin{equation} S_A= \frac{1}{3}\log\left[L \sin\left(\frac{\pi L_F}{L}\right)\sin\left(\frac{\pi L_A}{L}\right)\right] +\text{const.}\end{equation}

Essentially the same result has also been obtained via a much more complicated direct computation with Toeplitz matrices, albeit under the guise of spin XY chains\cite{its2005,its2007,its2009}.

\subsection{Duality between a $2$-band topological insulator and a bilayer Fermi gas}
We consider the 2-d two-band model with the Hamiltonian
\begin{eqnarray}
H=\sum_{\bf k}f_{\bf k}^\dagger {\bf \sigma\cdot d(k)}f_{\bf k}\label{Htwoband}
\end{eqnarray}
with spatial periodic boundary conditions. Here $\sigma$ are the Pauli matrices, and ${\bf d(k)}$ is a 3-d vector in momentum space which tends to a continuum in the thermodynamic limit. $f_{\bf k}$ is a two-component spinor in spin/pseudospin space. The single particle Hamiltonian has energy eigenvalues $E_\pm({\bf k})=\pm \left|{\bf d(k)}\right|$. If the vector ${\bf d(k)}$ is nonvanishing in the Brillouin zone, the system is an insulator with one band occupied at each ${\bf k}$.

We consider a real space partitioning of the system into parts $A$ and $B$. The single-particle operators $P$ and $R$ defined in Sec. \ref{sec:general} are
\begin{eqnarray}
R&=&\sum_{i\in A,\alpha=\uparrow,\downarrow}\ket{i,\alpha}\bra{i,\alpha}\nonumber\\
P&=&\sum_{\bf k, \alpha\beta}\frac{\delta_{\alpha\beta}-{\bf \sigma_{\alpha\beta}\cdot \hat{d}(k)}}{2}\ket{{\bf k},\alpha}\bra{{\bf k},\beta}
\end{eqnarray}

The dual system is defined by exchanging the roles of the real space coordinate $i$ and momentum ${\bf k}$. With periodic boundary condition in both directions, $i$ and ${\bf k}$ both take lattice values on a 2-d torus. After this re-identification, we obtain
\begin{eqnarray}
\tilde{R}&=&\sum_{i, \alpha\beta}\frac{\delta_{\alpha\beta}-{\bf \sigma_{\alpha\beta}\cdot \hat{d}(r_i)}}{2}\ket{i,\alpha}\bra{i,\beta}\nonumber\\
\tilde{P}&=&\sum_{{\bf k}\in A,\alpha=\uparrow,\downarrow}\ket{{\bf k},\alpha}\bra{{\bf k},\alpha}
\label{RPtilde}
\end{eqnarray}
Now $\tilde{R}$ is a real space projection to the spin direction $-{\bf \hat{d}(r_i)}$, as explored in for instance Refs. \onlinecite{doucot2008} and \onlinecite{puspus2014}, and $\tilde{P}$ is a projection to a momentum space region $A$. It can be interpreted as a projection to the Fermi sea for a Fermi gas with a spin rotation-symmetric Hamiltonian.
We can rewrite it as
\begin{eqnarray}
\tilde{P}&=&\sum_{{\bf k} ,\alpha=\uparrow,\downarrow}\theta\left(-\epsilon({\bf k})\right)\ket{{\bf k},\alpha}\bra{{\bf k},\alpha}
\end{eqnarray}
where $\epsilon({\bf k})$ is an energy dispersion such that $\epsilon({\bf k})<0$ only in region $A$. Its corresponding Hamiltonian is $H_{\rm dual}=\sum_{{\bf k},\sigma}\epsilon({\bf k})f_{{\bf k}\sigma}^\dagger f_{{\bf k}\sigma}$, which can also be written in real space as  
\begin{eqnarray}
H_{\rm dual}=\sum_{i,j,\sigma}f_{i\sigma}^\dagger t_{ij}f_{j\sigma}
\end{eqnarray}
as long as $A$ has a smooth boundary, with $t_{ij}$ being the Fourier transform of $\epsilon({\bf k})$.

To obtain a clearer physical picture of the dual system, one can perform a unitary rotation to a basis where $\tilde{R}$ is diagonal, so that $\tilde{R}$ can be reinterpreted as a real space projection. This can be inplemented through
\begin{equation}
U_i\frac{1-{\bf \sigma \cdot \hat{d}(r_i)}}{2}U_i^\dagger =\frac{1-\sigma_3}2\end{equation}
which rotates $\hat{\bf d}({\bf r_i})$ to the north pole on the Bloch sphere. 
In this new basis, the $H_{dual}$ and $\tilde R$ becomes
\begin{eqnarray}
H_{\rm dual}=\sum_{i,j,\alpha\beta}t_{ij}f_i^\dagger U_i^\dagger U_jf_j,\; \tilde{R}=\sum_{i}\ket{i,\downarrow}\bra{i,\downarrow}
\label{Rnew2}
\end{eqnarray}

In summary, the two bands in the original system have been transformed to two layers with a partition between them implemented by $\tilde R$. The Hamiltonian of the dual system describes fermion hopping in a non-Abelian background gauge field, with $e^{iA_{ij}}=U_i^\dagger U_j$, playing the role of a an $SU(2)$ gauge connection on the lattice which rotates among the two layers. When ${\bf d}({\bf k})$ in the original system is topologically nontrivial, the ES of both systems are identical and contain gapless chiral edge modes\cite{turner2009,qi2012}. In the dual system, the nontrivial ES results from the entangling of the two layers by the non-Abelian gauge field, even though the gauge field strength of a rotation is obviously trivial. In the next subsection, we will provide more a physical illustration of how the gapless ES arises in the dual system.

\subsubsection{Physical interpretation of the gapless ES  in the dual system}
\label{physical}

\begin{figure}[H]
\begin{minipage}{0.99\linewidth}
\includegraphics[scale=0.3]{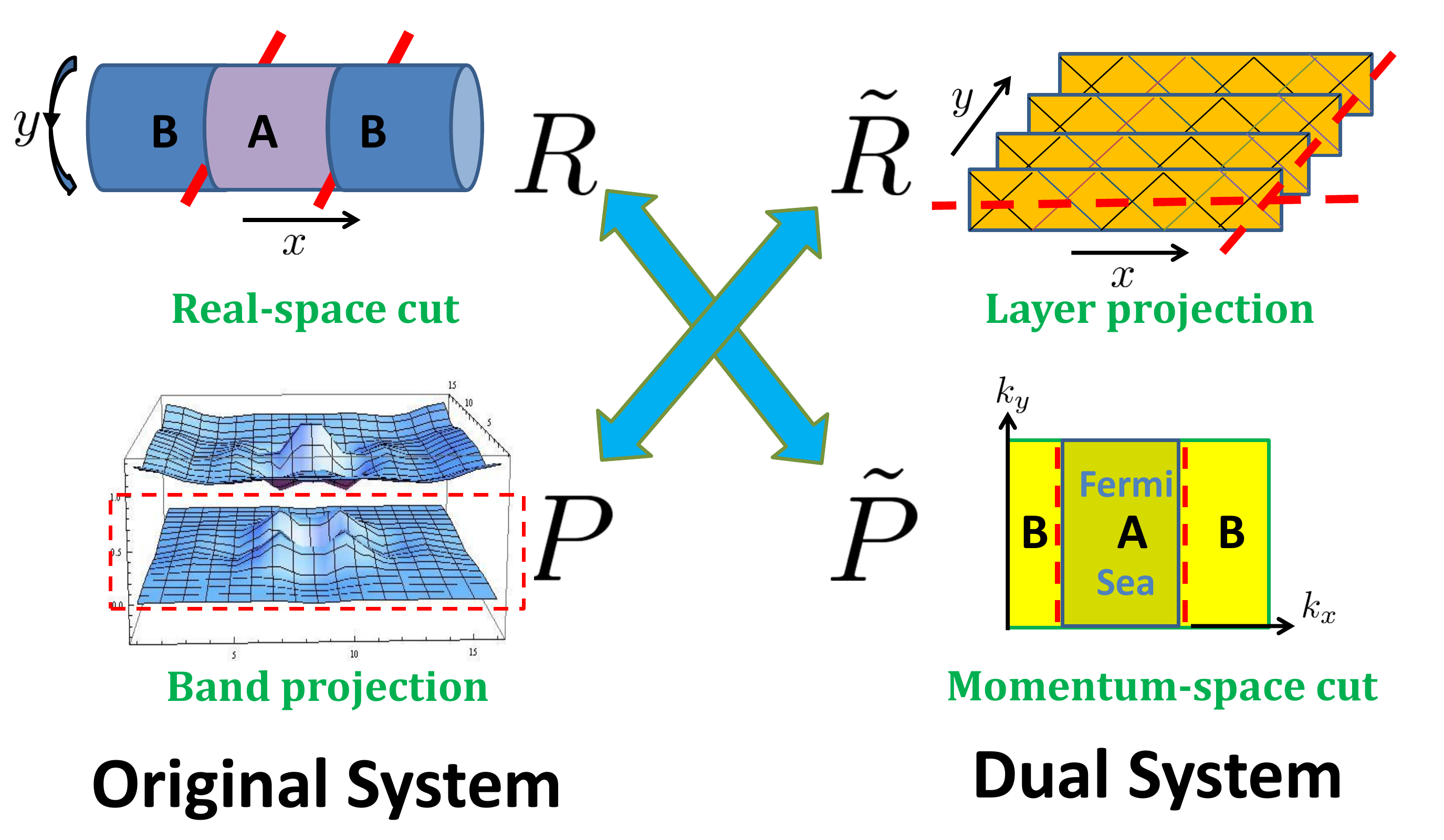}
\end{minipage}
\caption{ (Color Online) The original system (left) is a 2-band insulator. $R$ traces out region B and $P$ projects onto the occupied band. The dual system consists of decoupled bilayer 1-D chains with an $SU(2)$ gauge field coupling their two layers. $\tilde R$ now projects onto the lower layer while $\tilde P$ projects onto the occupied Fermi Sea of the same shape as region $A$ in the original system.}
\label {realkspace2}\end{figure}

In the previous example, the original system is a 2-d Chern insulator with chiral edge states and partitioned in real space. It has the same ES as a dual bilayer Fermi gas system in a background non-Abelian gauge field. Here, we shall  understand the nature of the gapless ES in the dual system more physically. For concreteness, we consider a partitioning of the original system which preserves translation symmetry in the $y$-direction, as shown in Fig. \ref{realkspace2}.
When its Chern number 
\begin{eqnarray}
C=\frac1{4\pi}\int d^2{\bf k}\hat{\bf d}({\bf k})\cdot \frac{\partial \hat{\bf d}({\bf k})}{\partial k_x}\times\frac{\partial \hat{\bf d}({\bf k})}{\partial k_y}\label{Chern2band}\end{eqnarray}
is nonzero, its ES contains gapless edge states \cite{turner2009,qi2012} with chiral dispersion in $k_y$ localized at the boundary between regions $A$ and $B$ .

The dual system thus have its Fermi surface along the $k_y$ direction, and its energy dispersion can be chosen to be 1-d: $\epsilon({\bf k})=\epsilon(k_x)$. In real space, the system consists of $L_y$ decoupled 1-d (2-layer) fermion chains, each coupled to a different external $SU(2)$ gauge field. From Eq. \ref{Rnew2}, they each have a real space partition between their two layers, as depicted in Fig. \ref{realkspace2}. 

To intuitively understand why the $SU(2)$ gauge field can give rise to a gapless ES, we perform a thought experiment of adiabatically switching off its interlayer coupling of the dual system by . Mathematically, we deform the hopping matrix element $t_{ij}U_i^\dagger U_j$ according to 
\begin{eqnarray}
e^{iA_{ij}}&=&U_i^\dagger U_j=\left(\begin{array}{cc}u_{ij}&v_{ij}\\-v_{ij}^*&u_{ij}^*\end{array}\right)\rightarrow \left(\begin{array}{cc}u_{ij}&\eta v_{ij}\\-
\eta v_{ij}^*&u_{ij}^*\end{array}\right)\notag\\
\end{eqnarray}
with $\eta\in(0,1]$ a real interpolation parameter.

When the two layers are decoupled in the limit $\eta \rightarrow 0$, the $SU(2)$ gauge field reduces to a $U(1)$ gauge field $u_{ij}=e^{\pm i a_{ij}=\left[e^{\pm iA_{ij}}\right]_{11}}$ for each layer. Although the $SU(2)$ gauge field $U^\dagger_i U_j$ is a pure gauge with zero field strength on the whole, the $U(1)$ gauge field $u_{ij}$ has a nonzero gauge curvature which actually corresponds to the Berry curvature in the original system\cite{bruno2004}, i.e. $a_{ij}=-\frac1{2(1+\hat d_3)}(\hat d_2 \nabla \hat d_1-\hat d_1 \nabla \hat d_2)$.

In the dual system of decoupled 1-d chains, the only gauge-invariant quantity of $a_{ij}$ is the flux along each 1-d line, labeled by its $y$ coordinate. In the continuum limit, this flux is $\Phi(y)=\int dxa_x(x,y)$, and the Chern number of the Berry's phase gauge field from Eq. (\ref{Chern2band})
is determined by the winding number of $\Phi(y)$ via 

\begin{equation}
C=\frac1{2\pi}\int\partial_y\Phi(y) dy
\end{equation}

\begin{figure}[H]
\begin{minipage}{0.99\linewidth}
\includegraphics[scale=0.5]{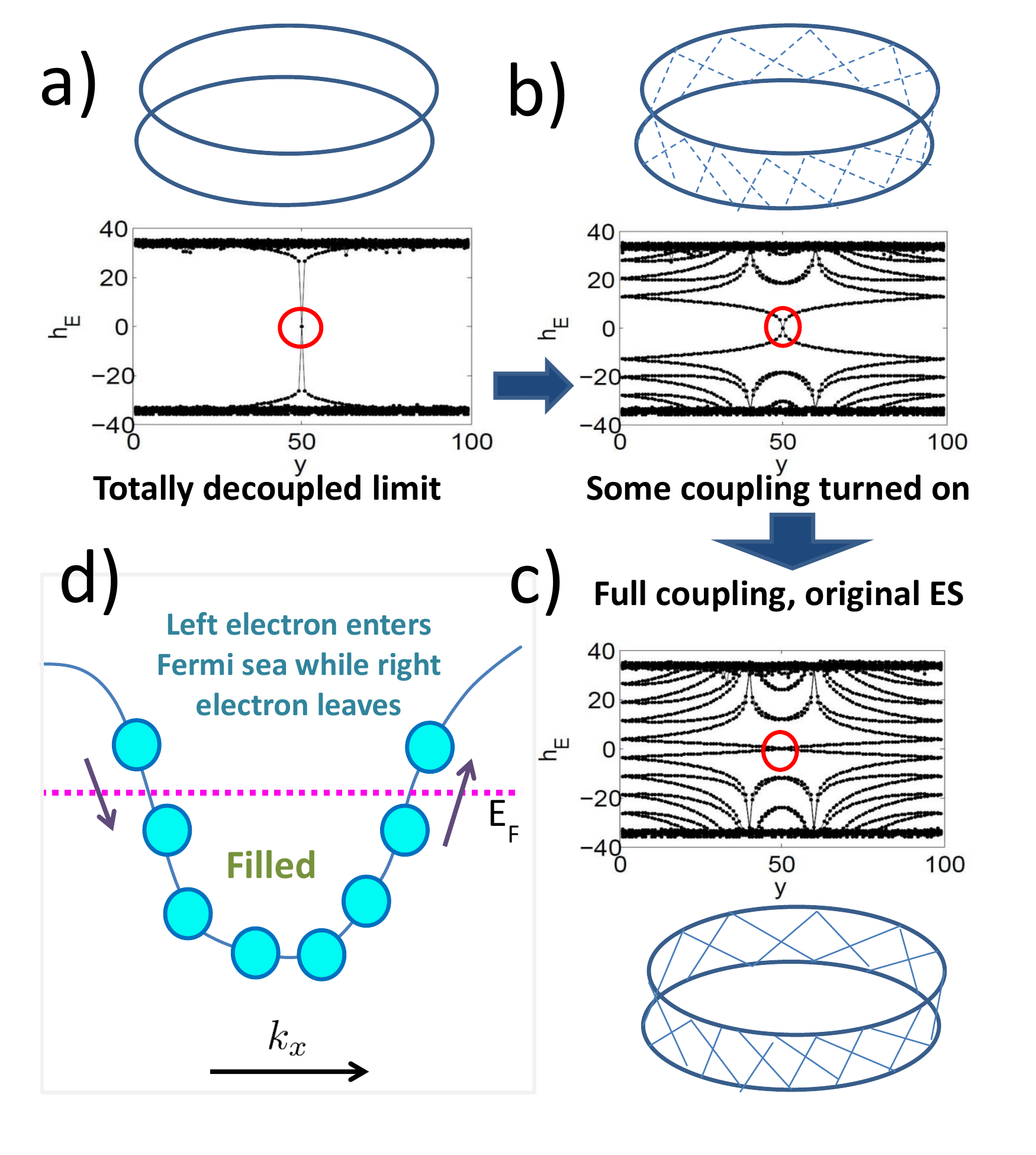}
\end{minipage}
\caption{(Color Online) a) to c) The interlayer ES and illustrations of a periodic 1-D chain in the dual system as a function of $y\in [1,L_y]$, $L_y=100$ with interpolation parameters $\eta=10^{-7},10^{-2},1$ respectively. Its gauge field taken to be that of a Dirac hamiltonian\cite{qi2008} with $m=0.8$. In the decoupled limit of a), there is negligible EE for most values of $y$ except at a degenerate point when $y=L_y/2$. The ES approaches that of the original system as interlayer coupling is restored. . d) As electrons shift in $k_x$, one enters the Fermi sea while another leaves. A level-crossing is inevitable.}
\label {interpol}\end{figure}

Therefore in the decoupled limit, we can interpret each decoupled 1-d chain layer as a 1-d Fermi gas on a ring threaded with flux $\pm\Phi(y)$. When $C\neq 0$, $\Phi(y)$ has a winding number of $C$ along the $y$ direction, so a spectral flow occurs as the parameter $y$ is varied. Specifically, as $y$ goes around a complete cycle, $C$ electrons states will be pumped\cite{king1993,coh2009,soluyanov2011} from the left to the right Fermi point, as illustrated in the inset of Fig. \ref{interpol}.
Since the Chern number of the other layer is equal and opposite, this pumping implies that an equal and opposite level crossing occurs at the Fermi surface of either layer as $y$ is varied.

When $\eta$ is gradually switched on, the two layers become coupled. When $\eta=1$, we recover the original system with an $SU(2)$ gauge field of zero field strength, and hence identical energy spectrum for each $y$. This means that with the introduction of interlayer coupling from $\eta$, there is no level crossing of the \emph{energy} spectrum. Since the level crossing can only be removed by coupling the two layers, the level crossing must be inherited by the ES for nonzero $\eta$. This is similar to the situation of inter-edge tunneling between two Quantum Hall states, where coupling the left and right moving edge states induces an energy gap, but the chiral dispersion is inherited by the ES between left- and right-movers\cite{turner2009,fidkowski2010,qi2012}. We have verified this physical argument numerically, as shown in Fig. \ref{interpol}, where the ES between the two layers for any $\eta\in (0,1]$ indeed has a level crossing as a function of $y$.

\subsection{Generalization to generic multi-band models}

The discussions in the last two subsections can be straightforwardly generalized to generic multiband systems. A general $N$-band system (which may be an insulator or a metal) with a real space partition is dual to a $N$-flavor Fermi gas coupled to an $SU(N)$ gauge field with entanglement partition determined by the Hamiltonian of the original system. An insulator Hamiltonian is dual to a interlayer partition, while a metal Hamiltonian is dual to some combination of layer and real space partitions.

More explicitly, the Hamiltonian of a translation invariant $N$-band system can be generically written as
\begin{eqnarray}
H=\sum_{\bf k}f_{\bf k}^\dagger h_{\bf k}f_{\bf k}
\end{eqnarray}
with $h_{\bf k}$ an $N\times N$ matrix and $f_{\bf k}$ an $N$-component spinor of the annihilation operators. Denoting the eigenstates and eigenenergies of $h_{\bf k}$ by $\ket{{\bf k},n}$ and $\epsilon_{n{\bf k}}$, the momentum and real space projection operators to region $A$ are
\begin{eqnarray}
P=\sum_{n,{\bf k}}\theta\left(-\epsilon_{n{\bf k}}\right)\ket{{\bf k},n}\bra{{\bf k},n},~R=\sum_{i\in A,\alpha}\ket{i\alpha}\bra{i\alpha}
\label{poriginal}
\end{eqnarray}
The dual system is defined like before, by exchanging real space coordinate ${\bf r}_i$ and momentum ${\bf k}$ and thus the roles of $P$ and $R$:
\begin{eqnarray}
\tilde{R}=\sum_{n,{\bf r}_i}\theta\left(-\epsilon_{n{\bf r}_i}\right)\ket{{\bf r}_i,n}\bra{{\bf r}_i,n},~\tilde{P}=\sum_{{\bf k}\in A,\alpha}\ket{{\bf k}\alpha}\bra{{\bf k}\alpha}
\end{eqnarray}
In the dual system, the band index $n$ is interpreted as a layer (or flavor) index, so that $\tilde{R}=P$ is a real space partition that projects onto layers corresponding to the original occupied bands. If the original system is a metal, some layers will also have additional real space partitions with boundaries corresponding to the original Fermi surface. We can again define a dual Hamiltonian
\begin{eqnarray}
H_{\rm dual}=\sum_{{\bf k},\alpha}\tilde{\epsilon}_{\bf k}f_{{\bf k}\alpha}^\dagger f_{{\bf k}\alpha}=\sum_{i,j,\alpha}t_{ij}f_{i\alpha}^\dagger f_{j\alpha}
\end{eqnarray}
such that its energy dispersion $\tilde{\epsilon}_{\bf k}<0$ in region $A$, and $t_{ij}$ denoting its Fourier transform.
In terms of original bands $\ket n$,
\begin{eqnarray}
H_{\rm dual}&=&\sum_{i,j,n,m}t_{ij}f_{in}^\dagger \left[U_i^\dagger U_j\right]_{nm}f_{jm}
\end{eqnarray}
where $U_{i}^{\alpha n}=\left\langle{\bf r}_i \alpha|{\bf r}_i n\right\rangle$.
Now, the dual system is a free Fermi gas coupled to a curvatureless $SU(N)$ gauge field $e^{iA_{ij}}=U_i^\dagger U_j$. When the original system has a gapless ES due to nontrivial topology, the identical ES of its dual system has a corresponding level crossing in real space due to the nontrivial gauge connection that entangles its different layers. 

In general, nontrivial (momentum space) band topology of the multi-band system will always be encoded in the real-space multi-component spin texture of the dual system. Thus one can, in principle, uncover analogous dualities involving systems with more general topological indices, i.e. the $Z_2$ index. That said, such generalizations must be made with care because the entanglement cut, which corresponds to the dual Fermi surface, may break the symmetry that protects the topological state in the first place. A detailed analysis of the position-momentum duality of generic symmetry protected topological states will be the subject of futher work, which may lead to interesting new physical pictures far beyond that in Sect. \ref{physical}.  


\subsection{The high-temperature limit and its relation to Wannier functions}

At nonzero temperature $T$, the eigenvalues of the equal-time correlation matrix $P$ follow the Fermi distribution $\frac1{e^{\beta E_{n{\bf k}}}+1}$, which has has a smeared momentum cutoff. Hence $P$ is no longer a projector. Therefore $Spec(RPR)\neq Spec(PRP)$, and 
$\tilde P \tilde R\tilde P$ can no longer be interpreted as a dual entanglement operator. 

Instead, it acquires an interesting interpretation in the $T\rightarrow \infty$ limit, where $P$ and hence its dual $\tilde R$ becomes linear in the position $r$. Restricting ourselves to the 1-D case, this can be seen by expanding the eigenvalues about $r=0$: $(1+e^{r/T})^{-1}\approx \frac{1}{2}\left(1-\frac{r}{2T}\right)$. Hence $\tilde R$ is essentially linear in $R$, i.e. a bona-fide position operator. $\tilde P\tilde R \tilde P$, which is now a position operator projected onto the occupied subspace, is nothing but a Wannier operator\cite{kivelson1982,yu2011}. Hence we have an identification of the Wannier polarization of a band insulator with the ES of its dual Fermi-liquid system in the high $T$ limit.

\section{Conclusion and discussion}

We have shown how the duality between the occupied-band projector $P$ and the real space partition $R$ in free fermion systems 
allows us to construct two very different systems with identical entanglement spectrum. With that, we show that a multi-flavored Fermi gas coupled to an external $SU(N)$ gauge field with trivial field strength can have a gapless ES between different flavors, with a dual system being a topological insulator partitioned in real space. As this example illustrates, the duality is helpful in understanding the ES when the dual system affords a more intuitive understanding. Another example of this duality is the EE formula of a simple Fermi liquid in $d$ dimensions $S=\frac{L^{d-1}\log L}{12(2\pi)^{d-1}}\int dA_x \int dA_k\left|n_{\bf x}\cdot n_{\bf k}\right|$, which shows that the real and momentum space integrals across the boundary $\int dA_x$ and the Fermi surface $\int dA_k$ play completely symmetric roles.

Since the position-momentum duality fundamentally rests on the invariance of the  eigenvalue spectra of $C$ under interchange of $P$ and $R$ projectors\cite{huang2012}, it holds for any basis-independent function of $C$. As such, the duality can be extended to other physical quantities, for instance the particle-number fluctuation $(\Delta n)^2 = Tr C(1-C)=Tr[PRP-PRPRP]$ in region $A$, which is often studied in quantum noise\cite{klich2009}. An open question is whether any form of such duality still holds when interaction is introduced to the system. 

\section{Acknowledgements}
This work is supported by a scholarship from the Agency of Science, Technology and Research of Singapore (C. H. L.) and the National Science Foundation through the grant No. DMR-1151786 (X. L. Q).
Research at Perimeter Institute is supported by the Government of Canada through Industry Canada and by the Province of Ontario through the Ministry of Research and Innovation. (P.Y.)

\bibliography{TI,entanglement}

\end{document}